\documentclass[aps,10pt,reprint,twocolumn]{revtex4}

\usepackage{epsfig}
\usepackage{graphicx}
\usepackage{dcolumn}
\usepackage{amsmath}
\usepackage{latexsym}
\usepackage{color}
\usepackage{subfig}
\usepackage{mathtools}
\usepackage{footnote}
\usepackage{hyperref}
\hypersetup{
    colorlinks=true,
    linkcolor=blue,
    citecolor=blue,
    filecolor=blue,      
    urlcolor=blue,
    hyperfootnotes=true
}

\begin{document}
\title{Born-Infeld corrections to holographic transport coefficients with spatially modulated chemical potential}  

\author{Ankur Srivastav\footnote{
\href{mailto:ankursrivastav@bose.res.in}{ankursrivastav@bose.res.in}}, Sunandan Gangopadhyay\footnote{\href{mailto:sunandan.gangopadhyay@gmail.com}{sunandan.gangopadhyay@gmail.com}\\       \href{mailto:sunandan.gangopadhyay@bose.res.in}{sunandan.gangopadhyay@bose.res.in}}}
\affiliation{Department of Astrophysics and High Energy Physics, S. N. Bose National Centre for Basic Sciences,\\ Block-JD, Sector-III, Salt Lake City,\\ Kolkata 700106, India}

\author{Ashis Saha}
\affiliation{Department of Physics, University of Kalyani,\\ Kalyani 741235, India}

\begin{abstract}
\noindent In this work, we have analytically computed the holographic transport coefficients for $(2+1)$-dimensional strongly coupled field theories, placed in a spatially modulated chemical potential along the $x$-direction, in the presence of Born-Infeld electrodynamics. We have obtained these coefficients in terms of the black hole horizon data only. We have observed that the Born-Infeld parameter affects these coefficients in a non-trivial way. We have, then, used these results to further study a holographic model for ultra-clean graphene near Dirac point. This is a two current model with an inhomogeneous holographic lattice.  

\end{abstract}

\maketitle


\section{Introduction}
\noindent  Gauge/Gravity duality has been at the forefront in our understanding of strongly coupled condensed matter systems for the last two decades. It has shed light on various universal properties in such systems which were otherwise less tractable. This duality has been applied extensively to study the unconventional superconductors and superfluids \cite{hhh,hhh1,sah,cph,gr1,gr2, gr3,sg1,gg1,js,rbsg,gg2,gg3,pmchl, Ankur, Ankur1, Ankur2, Ankur3}. For such strongly coupled systems, understanding the transport properties is one of the important problems in the gauge/gravity duality. It is partly because the DC response to such holographic theories is generically infinite due to translational invariance at the holographic boundary \cite{sah,cph}. However, it has been shown that one may explicitly break this translational invariance, and thus providing a mechanism for momentum relaxation, by explicitly constructing holographic lattices. Many such dual models having various mechanisms for momentum relaxations have been studied in recent years \cite{HST,YL,Donos1, Donos2, Donos3, Donos4, Donos5, Matteo, Niko1}. 
In this work, we have considered the possibility of momentum relaxation due to the presence of spatially modulated chemical potential in the $x$-direction at the boundary theory as shown in some previous works \cite{CLS,GTHJES, Donos6}. Models with such a spatially modulated chemical potential in one direction at the AdS boundary are known as inhomogeneous holographic lattice models.\\
\noindent Holographic transport coefficients have been obtained for various dual models before. Especially, it has been shown that these transport coefficients can be given entirely in terms of the blackhole horizon data only \cite{Donos1, Donos2, Donos3, Donos4, Donos5}. For a homogeneous holographic lattice model, a number of results have been obtained in recent years. Such homogeneous lattices are usually built using axion fields in the bulk theory. However, relatively less results are available for an inhomogeneous holographic lattice model. Such inhomogeneous lattices in the gauge/gravity duality may be conceptualized with the the inclusion of a spatially modulated chemical potential in a certain direction.  \\
\noindent Holographic transport coefficients for such an inhomogeneous holographic lattice have been analytically studied in \cite{Donos6}. However, the non-linear extension to its gauge sector has been missing from the literature. So we have, first, studied the role of Born-Infeld electrodynamics in this model. We have analytically obtained the Born-Infeld corrected expressions for the thermoelectric DC response to this inhomogeneous holographic lattice model of the strongly coupled system. The choice of this particular non-linear electrodynamics has been made because it is of interest to the holographic community for the following reasons. Firstly, it comes naturally from the low energy limit of string theory and deals with the self-energy problem of Maxwell theory. A further comment is in order now. In string theory, Born-Infeld (BI) correction to Maxwell theory has a natural interpretation where BI parameter, $b$, is like the tension of the string. The corrections around $b \rightarrow 0$ should be considered as stringy corrections which modify not only the gauge sector but also the gravity sector \cite{CVJ,aat}. A way to incorporate these stringy corrections is to add higher derivative terms to the gravity action. In particular, adding such corrections to the gravity action would lead to modifications to the usual Einstein field equations. A noteworthy analysis of holographic transport coefficients in the presence of Gauss-Bonnet gravity and more general higher derivative theories of gravity is given in \cite{Donos7}.\\ 
In this work, however, we have only considered the BI correction in the gauge sector of our model similar to other studies in holography \cite{Ankur2, Wu} . Also, such an approximation is suitable for analytic hold of calculations which would otherwise be very involved and obscure simple corrections to the holographic transport coefficients. Second, it is the only non-linear theory of electrodynamics that enjoys dual symmetry of the Maxwell theory \cite{MB,MBLI,Dirac,rbsg,gg3,Wu}. We have found that the presence of Born-Infeld electrodynamics affects the thermoelectric DC response in a non-trivial way. It has been observed that Lorentz factors obtained in this analysis are neither constant nor equal and thus shows a violation to the Wiedemann-Franz law, which asserts that the Lorentz factors must be constant in the weakly coupled system. Hence it is indicative of the strong coupling at the boundary field theory.\\
\noindent We have then used these BI corrected results to, further, built a two current model for ultra-clean Graphene near Dirac point as shown in \cite{seo, Rogatko1, Rogatko2}. It has been argued in \cite{seo} that these two currents can be understood as separate contributions of electrons and holes near the Dirac point. This second hidden current then improves the single current model so as to explain the experimental results. \\
\noindent We have, here, considered two Born-Infeld currents in the bulk theory in a planar-Schwarzschild $AdS_{3+1}$ blackhole spacetime with spatially modulated chemical potential in the $x$-direction, unlike \cite{seo}, where a homogeneous holographic lattice is built using axion fields. It is to be noted that the holographic model in \cite{seo} worked under the assumption that charges of the two currents are proportional to each other. Hence, we have considered  an ansatz about time components of the gauge fields at the horizon of the black hole and also about the BI parameters of the two currents so that we get the charges associated with both currents to be proportional to each other. Under these assumptions, we have obtained the final expressions for the thermoelectric DC response which may be directly compared with the thermoelectric DC response of the single current model in terms of the proportionality constant of the charges in two current model. We have kept these results general so that the low and high temperature behaviours could be obtained very easily as shown in \cite{Donos6}. Also, results for the inhomogeneous holographic lattice model with two Maxwell currents, which have not been addressed before, could be readily obtained by considering BI parameters to be zero in the final expressions for the thermoelectric DC response. \\   
\noindent This paper is organised as follows. We start with setting up the holographic model in section (\hyperlink{sec2}{II}) in a static black hole background in a planar-Schwarzschild $AdS_{3+1}$ spacetime. Then in section (\hyperlink{sec3}{III}), we have analytically calculated the holographic transport coefficients for the model. Section (\hyperlink{sec4}{IV}) deals with the application of these results for a two current model built to understand ultra-clean Graphene in a similar way as proposed in \cite{seo}. We have made final comments and concluding remarks on the results in the last section (\hyperlink{sec5}{V}) of this paper.  
 \section{The Holographic Model}
\noindent \hypertarget{sec2}{We} start with writing the following action consisting of gravity and gauge sector,
\begin{eqnarray}
\mathcal{S} = \int d^4x \sqrt{-g} \bigg(R+6+\mathcal{L}_{BI}\bigg)
\label{Action}
\end{eqnarray}
where the cosmological constant, $\Lambda =-3$, and we have set the AdS radius to be unity, and $16\pi G=1$. Here the gauge sector in the action is given by $\mathcal{L}_{BI}$, which is the Lagrangian density for Born-Infeld electrodynamics \cite{rbsg,gg3,Wu}. It is given by the following expression,
\begin{eqnarray}
\mathcal{L}_{BI} =\dfrac{1}{b}  \bigg(1-\sqrt{1+\dfrac{b}{2}F^2}\bigg)
\label{BI_Lagrangian}
\end{eqnarray}
where $F^2=F_{\mu\nu}F^{\mu\nu}$ and the Faraday tensor is given by $F_{\mu\nu}=\partial_{[\mu}A_{\nu]}$.\\
\noindent The equations of motion associated with the action (\ref{Action}) are given by,
\begin{eqnarray}
 E_{\mu\nu} \equiv R_{\mu\nu}+3g_{\mu\nu}-\dfrac{1}{2}T_{\mu\nu}=0~~
\label{EOM_I} \\
\nabla_{\mu}\bigg(\dfrac{F^{\mu\nu}}{\sqrt{1+\dfrac{b}{2}F^2}}\bigg)=0~.
 \label{EOM_II}
\end{eqnarray}
Here $E_{\mu\nu}$ is the Einstein tensor and $T_{\mu\nu}$ is the energy-momentum tensor associated with $\mathcal{L}_{BI}$, which is given by \cite{Wu}, 
\begin{eqnarray}
T_{\mu\nu} = g_{\mu\nu}\mathcal{L}_{BI}+\dfrac{F_{\mu\rho}F_{\nu}^{\rho}}{\sqrt{1+\dfrac{b}{2}F^2}}~.
 \label{EM tensor}
\end{eqnarray}
We now consider the model given by action (\ref{Action}) in the planar-Schwarzschild AdS blackhole geometry, which is given by the following metric,
\begin{eqnarray}
 ds^{2}= -U(r) dt^2+\dfrac{dr^2}{U(r)}+\Sigma(r)(dx^2+dy^2)~.
 \label{metric}
\end{eqnarray}
In this geometry, the blackhole horizon is at $r=0$ such that $U(r=0) = 0$. We now introduce the spatially modulated chemical potential in the boundary theory along the $x$-direction, $\mu(x)$, given by \cite{CLS,GTHJES, Donos6},
\begin{eqnarray}
\mu(x)=\mu_0 + \bar\mu(x)
 \label{Mod Chem}
\end{eqnarray}
such that $\bar\mu(x)=\bar\mu(x+L)$, with period L. Here $\mu_0$ is the constant part of the chemical potential. Such a modulated chemical potential can be introduced on the bulk side in the dual model by considering the following choice for the bulk gauge-field, 
\begin{eqnarray}
A_{\mu}=(a_t(r,x),0,0,0)~.
\label{Gauge Field}
\end{eqnarray}
\noindent Regularity condition for the gauge-field at the blackhole horizon implies that $a_t(0,x)=0$. Apart from it, we now assume the following near horizon expansions for gauge-field and the metric coefficients as mentioned in \cite{Donos6},
\begin{eqnarray}
a_t(r,x)=ra_t^{(0)}(x)+r^2a_t^{(1)}(x)+\mathcal{O}(r^3)
\label{Near_Horizon_Expansions1}\\
U(r)=4\pi Tr+U^{(1)}r^2+\mathcal{O}(r^3) ~~~~~~~~~\label{Near_Horizon_Expansions2}\\
\Sigma(r)=\Sigma^{(0)}+r\Sigma^{(1)}+\mathcal{O}(r^2)~.~~~~~~~~~~~
\label{Near_Horizon_Expansions3}
\end{eqnarray}
Here  $T$ is the Hawking temperature of the blackhole. In general, $\Sigma(r)$ should be replaced with $\Sigma(r,x)$ as it should be spatially modulated in the $x$-direction. However to keep the calculations simple, we have assumed $\Sigma(r)$ to be independent of $x$ as considered in \cite{seo}.\\
\noindent The electric current densities for the dual field theory at the boundary, are defined by the following expression,
\begin{eqnarray}
J^a \equiv \bigg( \dfrac{\sqrt{-g}F^{ar}}{\sqrt{1+\dfrac{b}{2}F^2}}\bigg)|_{r\rightarrow \infty}~.
\label{Current Densities}
\end{eqnarray}
Using eq.(\ref{Current Densities}), we may define total charge of the blackhole, 
\begin{eqnarray}
q \equiv \dfrac{1}{L} \int_{0}^{L} dx J^t
\label{BH Charge}
\end{eqnarray}
where the integral in $x$-direction is taken over a period of holographic lattice, $L$. From now onwards, we would be denoting integral over a lattice period in $x$-direction using the following notation,
 $$\dfrac{1}{L} \int_{0}^{L} dx \equiv \int~.$$ 

\section{The Transport Coefficients}
\noindent \hypertarget{sec3}{In} this section, we would be computing the holographic transport coefficients. We would obtain these coefficients, within the linear response of the system, for thermoelectric perturbations to the boundary field theory. To do so, we would be externally applying a constant electric field $E$, and a thermal gradient $\zeta \equiv -\dfrac{\nabla T}{T}$ to the boundary field theory, which could be realised in the bulk theory as some perturbations to the gauge fields. However, these perturbations would, then, demand certain perturbations in the background metric itself \cite{Donos6}. Within the linear response regime, we may obtain various thermoelectric coefficients defined by the following matrix equation, 
\begin{eqnarray}
\begin{pmatrix}
    J        \\
    \mathcal{Q}     
\end{pmatrix}
=
\begin{pmatrix}
    \sigma       &  \alpha T \\
    \bar \alpha T    & \bar \kappa T 
\end{pmatrix}
\begin{pmatrix}
    E      \\
    \zeta  
\end{pmatrix}~.
\label{Matrix}
\end{eqnarray}
In the above matrix, $\sigma$ is the electric conductivity, $\bar \kappa$ is the thermal conductivity while $\alpha$ and $\bar \alpha$ are known as thermoelectric conductivities. $J$ and $\mathcal{Q}$ are electric and heat currents respectively. In general, the following relation is known to hold in real materials, $$\bar \alpha = \alpha$$ which is known as the Onsager relation.\\  
\noindent Now, from the $t$-component of the gauge equations of motion, eq.(\ref{EOM_II}), we get,
$$\nabla_{r}\bigg(\dfrac{F^{r t}}{\sqrt{1+\dfrac{b}{2}F^2}}\bigg)=0.$$ 
This equation implies that charge, $q$, remains constant along the AdS direction, $r$, and hence, can be evaluated anywhere in the AdS direction including the horizon at $r=0$. Therefore, using the near horizon expansions given by eqs.(\ref{Near_Horizon_Expansions1}, \ref{Near_Horizon_Expansions2}, \ref{Near_Horizon_Expansions3}), we get the Born-Infeld corrected expression for charge in terms of the blackhole horizon data upto $\mathcal{O}(r)$ as,
\begin{eqnarray}
q=\int \Sigma^{(0)} a_t^{(0)} \bigg(1+\dfrac{b}{2}(a_t^{(0)})^2\bigg)~.
\label{Charge}
\end{eqnarray}
We now apply a time independent constant electric field $E$ at the boundary theory, which may be introduced on the bulk side as a linear in time perturbation in the gauge field. This could be realised from the following perturbations in the metric and the gauge fields \cite{Donos6}, 
\begin{eqnarray}
\delta ds^2 = \delta g_{tt} dt^2 + \delta g_{rr} dr^2 + \delta g_{xx} dx^2 + \delta g_{yy} dy^2 \nonumber \\
+ 2\delta g_{tr} dtdr+ 2\delta g_{tx} dtdx+2\delta g_{xr} dxdr ~~
\label{BH_Perturbations1}\\ 
 \delta A _{\mu}= (\delta a_t, \delta a_r, \delta a_x - Et, 0)~.~~~~~~~~~~~~~~~~~~~~
\label{BH_Perturbations2}
\end{eqnarray}
Notice that all these perturbations, except electric field term $Et$, are functions of both $x$ and $r$ coordinates. Also these perturbations are periodic in the $x$-direction with the same period, $L$, with which chemical potential is modulated in the $x$-direction. Just like the gauge field and metric coefficients given in eqs.(\ref{Near_Horizon_Expansions1}, \ref{Near_Horizon_Expansions2}, \ref{Near_Horizon_Expansions3}), we consider that these perturbations also have the following near horizon expansions,
$$\delta a_t=\delta a_t^{(0)}+\mathcal{O}(r)~;~~~~~~~~~~\delta a_r=\dfrac{1}{U}(\delta a_r^{(0)}+\mathcal{O}(r)) ~~$$
$$\delta a_x= \ln U(\delta a_x^{(0)}+\mathcal{O}(r))~;~~
\delta g_{tt}= U (\delta g_{tt}^{(0)}+\mathcal{O}(r)) ~~~$$
$$\delta g_{rr}= \dfrac{1}{U} (\delta g_{rr}^{(0)}+\mathcal{O}(r))~;~~~~
\delta g_{xx}= \delta g_{xx}^{(0)}+\mathcal{O}(r) ~~~~~~~~$$
$$\delta g_{yy}= \delta g_{yy}^{(0)}+\mathcal{O}(r)~;~~~~~~~~~~
\delta g_{tr}= \delta g_{tr}^{(0)}+\mathcal{O}(r) ~~~~~~~~$$
$$\delta g_{tx}= \delta g_{tx}^{(0)}+\mathcal{O}(r)~;~~~~~~~~~
\delta g_{xr}=  \dfrac{1}{U}(\delta g_{xr}^{(0)}+\mathcal{O}(r))~. 
$$
The factors of $U$ in some of the perturbations are appropriately chosen in such a way that the regularity conditions at the horizon look simple. This shall be apparent in the following subsection. 


\subsection{Regularity Conditions}

\noindent We need these perturbations to be regular at the horizon of the blackhole. Therefore, to discuss the regularity conditions for these perturbations at the blackhole horizon, we switch to ingoing Eddington-Finkelstein (EF)-coordinates given by, 
\begin{eqnarray}
v=t+\int dr \sqrt{\dfrac{g_{rr}}{g_{tt}}}~.
\label{EF_Coord}
\end{eqnarray}
For the background  metric given by eq.(\ref{metric}), near the blackhole horizon, this leads to $v \simeq  t+\dfrac{ln r}{4\pi T}$. Now considering the metric perturbations, $\delta ds^2$, in EF-coordinates and requiring these to be regular at the horizon, $r=0$, leads to the following constraints, 
\begin{eqnarray}
\delta g_{tt}^{(0)} + \delta g_{rr}^{(0)} -2 \delta g_{tr}^{(0)} = 0 
\label{Constraint1} \\
\delta g_{xr}^{(0)} = \delta g_{tx}^{(0)}~.~~~~~~~~~~~~~~~
\label{Constraint2} 
\end{eqnarray}
The first constraint, eq.(\ref{Constraint1}), is due to regularity of the coefficient of $dr^2$ at the horizon, while the second constraint, eq.(\ref{Constraint2}), comes from requiring regularity of the coefficient of $drdx$. Coefficients of the other metric components do not lead to any further constraints on the perturbations at the leading order in $r$. Let us now analyse the gauge field perturbations in the similar way. Writing gauge field perturbations in the EF coordinates and demanding regularity at the blackhole horizon leads to following constraints on these perturbations,
\begin{eqnarray}
\delta a_r^{(0)} = \delta a_t^{(0)}~~~~
\label{Constraint3} \\
\delta a_x^{(0)} = - \dfrac{E}{4\pi T}~.
\label{Constraint4} 
\end{eqnarray}


\subsection{Gauge Currents}
\noindent We are now in a position to obtain the expressions for the gauge currents associated with the perturbations defined in eqs.(\ref{BH_Perturbations1}, \ref{BH_Perturbations2}). At first, we would compute the electric current, $J$, due to the applied electric field, $E$. From gauge equations of motion given by eq.(\ref{EOM_II}) along $r$ and $x$ directions, we find that 
$$\bigg( \dfrac{\sqrt{-g}F^{xr}}{\sqrt{1+\dfrac{b}{2}F^2}}\bigg) = constant$$
in both $r$ and $x$. So we may evaluate this expression anywhere along the AdS-direction, including the blackhole horizon at $r=0$. It is clear from eq.(\ref{Current Densities}) that the expression mentioned above gives the electric current density at the AdS boundary,
\begin{eqnarray}
J \equiv J^x = \bigg( \dfrac{\sqrt{-g}F^{xr}}{\sqrt{1+\dfrac{b}{2}F^2}}\bigg)|_{r\rightarrow \infty}~.
\label{electric current1}
\end{eqnarray}
To first order in the perturbations, the electric current density is given by,
\begin{eqnarray}
J  =  \bigg(1+\dfrac{b}{2}\{(\partial_r \delta a_t)^2+\dfrac{1}{\Sigma U}(\partial_x \delta a_t)^2\}\bigg)[U (\partial_x \delta a_r - \partial_r \delta a_x) \nonumber \\
+ (\partial_x  a_t) \delta g_{tr} - (\partial_r  a_t) \delta g_{tx}]~.~~~~~
\label{electric current2}
\end{eqnarray}
Now using the near horizon expansions for these perturbations mentioned above in the eq.(\ref{electric current2}), we obtain, at the leading order in $r$, the following expression for $J$,
\begin{eqnarray}
J  =  \bigg(1+\dfrac{b}{2} (a_t^{(0)})^2\bigg)(E + \partial_x \delta a_t^{(0)} - a_t^{(0)} \delta g_{tx}^{(0)})~.
\label{electric current3}
\end{eqnarray}
Next, we focus on the heat current associated with these perturbations. We start by noting that under reasonable symmetry considerations as mentioned in \cite{Donos6}, one may define the heat current, to first order in perturbations, by the following expression \footnote{It is well known that the gravity action must be supplemented with a boundary term, for well posed variational problem, and counter terms to cancel the divergences arising due to infinite volume of spacetime. With these boundary terms and counter terms in the action, a well defined boundary stress tensor for AdS spacetime was given in \cite{Bala, HSS}. For general class of gravity theories, it has been shown in \cite{HSL} that the radially conserved bulk current evaluated at the boundary matches with the thermal current on the boundary obtained using the boundary stress tensor. It has been further discussed in \cite{HSL} that since the boundary spacetime is conformally flat, no additional counter terms associated with the curvature of the boundary are needed.},
\begin{eqnarray}
\mathcal{Q} = 2 \sqrt{-g} \nabla^r K^x - a_t J
\label{Heat Current1}
\end{eqnarray}
where $K = (\dfrac{\partial}{\partial t}, 0, 0, 0)$ is the Killing vector. To first order in the perturbations, this would further simplify to the following expression,
\begin{eqnarray}
\mathcal{Q} = U^2\bigg(\partial_r\bigg(\dfrac{\delta g_{tx}}{U}\bigg) - \partial_x\bigg(\dfrac{\delta g_{tr}}{U}\bigg)\bigg) - a_t J~.
\label{Heat Current2}
\end{eqnarray}
Using the near horizon expansions for the perturbations along with leading order in $r$ expression for $J$ from eq.(\ref{electric current3}) in the eq.(\ref{Heat Current2}), we get the following result for heat current at leading order in $r$,
\begin{eqnarray}
\mathcal{Q} = -4\pi T \delta g_{tx}^{(0)} = constant~.
\label{Q_at_leading}
\end{eqnarray}
This implies that $\delta g_{tx}^{(0)}$ is a constant. Further, at subleading order in $r$, we also get the following constraint, 
\begin{eqnarray}
\partial_x(4\pi T\delta g_{tr}^{(0)}) +\delta g_{tx}^{(0)}\bigg(2 U^{(1)} - (a_t^{(0)})^2 \bigg(1+\dfrac{b}{2}(a_t^{(0)})^2\bigg)\bigg) ~\nonumber \\
+ a_t^{(0)}\bigg(1+\dfrac{b}{2} (a_t^{(0)})^2\bigg)(E + \partial_x \delta a_t^{(0)}) = 0~.~~~~~~~~~~~~~~~~~~~~
\label{Constraint5}
\end{eqnarray}
Now using the background (that is for unperturbed metric) Einstein equations ($E_{tt}$ and $E_{xx}$), we obtain the following condition that relates $U^{(1)}$ to the horizon data for the gauge field, that is $a_t^{(0)}$,
\begin{equation}
2 U^{(1)} - (a_t^{(0)})^2 \bigg(1+\dfrac{b}{2}(a_t^{(0)})^2\bigg) = 0~. 
\label{Constraint6}
\end{equation}
Substituting eq.(\ref{Constraint6}) in eq.(\ref{Constraint5}), we get the following condition, which consists of the blackhole horizon data only,
\begin{eqnarray}
\partial_x(4\pi T\delta g_{tr}^{(0)}) + a_t^{(0)}\bigg(1+\dfrac{b}{2} (a_t^{(0)})^2\bigg)(E + \partial_x \delta a_t^{(0)}) = 0~.~~
\label{Constraint7}
\end{eqnarray}
We would use eq.(\ref{electric current3}) to write eq.(\ref{Constraint7}) in a more useful way, as shown below,
\begin{eqnarray}
\partial_x(4\pi T\delta g_{tr}^{(0)}) + a_t^{(0)}\bigg(J + a_t^{(0)} \bigg(1+\dfrac{b}{2} (a_t^{(0)})^2\bigg)\delta g_{tx}^{(0)}\bigg)= 0~.~~
\label{Constraint8}
\end{eqnarray}
Integrating this expression over a period of $x$, we would get the following result, 
\begin{eqnarray}
J \int a_t^{(0)} = \dfrac{\mathcal{Q}}{4\pi T} \int (a_t^{(0)})^2 \bigg(1+\dfrac{b}{2} (a_t^{(0)})^2\bigg)~.
\label{JQ1}
\end{eqnarray}
In getting the above result, we have used eq.(\ref{Q_at_leading}) to replace $\delta g_{tx}^{(0)}$ in terms of $\mathcal{Q}$. Notice that the term involving $g_{tr}^{(0)}$ vanishes due to periodicity. Also, rewriting eq.(\ref{electric current3}) in the following form,
\begin{eqnarray}
\dfrac{J}{\bigg(1+\dfrac{b}{2} (a_t^{(0)})^2\bigg)}  = E + \partial_x \delta a_t^{(0)} + a_t^{(0)} \dfrac{\mathcal{Q}}{4\pi T} 
\label{electric current4}
\end{eqnarray}
and integrating it over a period of $x$, we find,
\begin{eqnarray}
J \int \dfrac{1}{\bigg(1+\dfrac{b}{2} (a_t^{(0)})^2\bigg)}  = E +  \dfrac{\mathcal{Q}}{4\pi T} \int a_t^{(0)}~.
\label{electric current5}
\end{eqnarray}
With some simple manipulations with eqs.(\ref{JQ1}, \ref{electric current5}), It is easy to obtain the following expressions for the DC responses due to constant electric field, $E$,
\begin{eqnarray}
\sigma \equiv \dfrac{J}{E} = \dfrac{1}{X_b} \bigg(1 + \dfrac{(\int a_t^{(0)})^2}{X}\bigg) \label{response1}\\
\bar \alpha \equiv \dfrac{\mathcal{Q}}{TE} = \dfrac{4\pi \int a_t^{(0)}}{X}~~~~~~~~~~~~
\label{response2}
\end{eqnarray}
where $X_b$ and $X$ are given by,
\begin{eqnarray}
X_b \equiv \int \dfrac{1}{\bigg(1+\dfrac{b}{2} (a_t^{(0)})^2\bigg)} ~~~~~~~~~~~~~~~~~~~~~~~~~~~~~~~~~~~~  \label{Xb_means}\\
X \equiv  \bigg[\int (a_t^{(0)})^2 \bigg(1+\dfrac{b}{2} (a_t^{(0)})^2\bigg)\bigg] X_b - \bigg(\int a_t^{(0)}\bigg)^2~.~~ 
\label{X_means}
\end{eqnarray}
To summarize, we have so far obtained the linear DC responses to the constant electric field, $E$, in our spatially modulated chemical potential holographic model. Now we shall obtain the linear DC response for this holographic model due to an applied temperature gradient, $\zeta$, at the AdS boundary. Again we shall apply this thermal gradient along $x$- direction, which may be realised with the following perturbations \cite{Donos6},
\begin{eqnarray}
\delta ds^2 = \delta g_{tt} dt^2 + \delta g_{rr} dr^2 + \delta g_{xx} dx^2 + \delta g_{yy} dy^2 \nonumber \\
+ 2\delta g_{tr} dtdr+ 2(\delta g_{tx} - tU\zeta) dtdx+2\delta g_{xr} dxdr
\label{BH_Perturbations3}\\ 
 \delta A _{\mu}= (\delta a_t, \delta a_r, \delta a_x + ta_t\zeta, 0)~.~~~~~~~~~~~~~~~~~~~
\label{BH_Perturbations4}
\end{eqnarray}
These perturbations are appropriately chosen so that time dependency drops out from the gauge currents. Regularity conditions for these perturbations at the horizon demands the following constraints,
\begin{eqnarray}
\delta g_{tt}^{(0)} + \delta g_{rr}^{(0)} -2 \delta g_{tr}^{(0)} = 0
\label{Constraint9} \\
\delta g_{xr}^{(0)} = \delta g_{tx}^{(0)}~~~~~~~~~~~~~~~~
\label{Constraint10}  \\
\delta g_{tx}^{(l)} = -\dfrac{\zeta}{4\pi T}~.~~~~~~~~~~~~
\label{Constraint11}
\end{eqnarray}
The new constraint involving $\delta g_{tx}^{(l)}$  is due to the regularity of the metric coefficient of $dtdx$. $\delta g_{tx}^{(l)}$ appears in the near horizon expansion of $\delta g_{tx}$ in the following way \cite{Donos1,Donos2,Donos3,Donos4,Donos5, Donos6},
$$\delta g_{tx}= \delta g_{tx}^{(0)}+ \delta g_{tx}^{(l)} (U \ln U)+\mathcal{O}(r)~.$$
Also, once again the regularity of the gauge perturbations at the blackhole horizon now leads to the following constraints, 
\begin{eqnarray}
\delta a_r^{(0)} = \delta a_t^{(0)}
\label{Constraint12} \\
\delta a_x^{(0)} = 0~.~~~
\label{Constraint13} 
\end{eqnarray}
Proceeding in a similar fashion as before, for these new perturbations, we get the following expressions for the gauge currents at leading order in $r$,
\begin{eqnarray}
J  =  \bigg(1+\dfrac{b}{2} (a_t^{(0)})^2\bigg)( \partial_x \delta a_t^{(0)} - a_t^{(0)} \delta g_{tx}^{(0)}) 
\label{electric current4} \\
\mathcal{Q} = -4\pi T \delta g_{tx}^{(0)}~.~~~~~~~~~~~~~~~~~~~~~~~~~~~~~~~
\label{Q_at_leading1} 
\end{eqnarray}
However, subleading order in $r$ expansion of the heat current now leads to the following constraint,
\begin{eqnarray}
\partial_x(4\pi T\delta g_{tr}^{(0)}) + a_t^{(0)}\bigg(1+\dfrac{b}{2} (a_t^{(0)})^2\bigg)(\partial_x \delta a_t^{(0)}) + 4\pi T \zeta= 0~.~~
\nonumber
\end{eqnarray}
Here we have utilised eq.(\ref{Constraint6}) in order to get the above equation in terms of the horizon data only. We could again write it in the following useful form using eq.(\ref{electric current4}),
\begin{eqnarray}
\partial_x(4\pi T\delta g_{tr}^{(0)}) + a_t^{(0)}\bigg(J + a_t^{(0)} \bigg(1+\dfrac{b}{2} (a_t^{(0)})^2\bigg)\delta g_{tx}^{(0)}\bigg) \nonumber \\
+ 4\pi T \zeta = 0. ~~
\label{Constraint15}
\end{eqnarray}
Integrating eq.(\ref{Constraint15}) over a period of $x$, we would get the following result, 
\begin{eqnarray}
J \int a_t^{(0)} + 4\pi T \zeta= \dfrac{\mathcal{Q}}{4\pi T} \int (a_t^{(0)})^2 \bigg(1+\dfrac{b}{2} (a_t^{(0)})^2\bigg)
\label{JQ2}
\end{eqnarray}
where we have used eq.(\ref{Q_at_leading1}) to replace $\delta g_{tx}^{(0)}$ in terms of $\mathcal{Q}$. Again using simple mathematical rearrangements of eqs.(\ref{electric current4}, \ref{Q_at_leading1}, \ref{JQ2}), we could easily obtain the following DC responses to the thermal gradient,
\begin{eqnarray}
\bar \kappa \equiv \dfrac{\mathcal{Q}}{T\zeta} = \dfrac{(4\pi)^2 T X_b}{X} \label{response3}\\
\alpha \equiv \dfrac{J}{T\zeta} = \dfrac{4\pi \int a_t^{(0)}}{X}~.
\label{response4}
\end{eqnarray}
Notice that we have found $\alpha = \bar \alpha$, which is the Onsager relation. We may also define the thermal heat current when all other currents are absent from the system, hence setting $J=0$, we get the following result, 
\begin{eqnarray}
 \kappa \equiv \dfrac{\mathcal{Q}}{T\zeta}\bigg|_{J=0} =  \dfrac{(4\pi)^2 T}{\int (a_{t}^{(0)})^2 \bigg(1+ \dfrac{b}{2}(a_{t}^{(0)})^2\bigg)}~.
 \label{heat current}
\end{eqnarray}
We may also obtain the Lorentz factors as below,
\begin{eqnarray}
 \bar l \equiv \dfrac{\bar \kappa}{\sigma T}= \dfrac{(4\pi)^2 X_b}{\int (a_{t}^{(0)})^2 \bigg(1+ \dfrac{b}{2}(a_{t}^{(0)})^2\bigg)} ~~ \label{Lorentz1}\\
l \equiv \dfrac{\kappa}{\sigma T} = \dfrac{\bar l X}{\int (a_{t}^{(0)})^2 \bigg(1+ \dfrac{b}{2}(a_{t}^{(0)})^2\bigg)}~.
\label{Lorentz2}
\end{eqnarray}
We note that these Lorentz factors, denoted by $\bar l$ and $l$, are neither constants nor equal to each other, which is indicative of the strongly coupled boundary theory.\\
\noindent To summarize, we have analytically computed non-trivial Born-Infeld corrections to all the thermoelectric DC responses for the holographic model with spatially modulated chemical potential. We have also given the Born-Infeld corrected expressions for the thermal conductivity in the absence on any other current as well as the expressions for the Lorentz factors. Next we would briefly discuss the thermal DC responses in a two current model, built with two independent Born-Infeld currents, for this spatially modulated chemical potential setting, which could be useful for the holographic modeling of the ultra-clean Graphene near charge neutrality point as shown in \cite{seo}.
\section{Two Current Model}
\noindent \hypertarget{sec4}{In} \cite{seo}, a two current holographic model for the ultra-clean Graphene near the Dirac point was presented. Under the assumption that the charges associated with both independent currents are proportional to each other, they have shown that their results are in agreement with the experiments. They have worked with a homogeneous holographic lattice model built with the axion fields. Here, we are now considering an inhomogeneous holographic lattice model built with the spatially modulated chemical potential and two independent Born-Infeld currents. In this section, we would use the results from the previous sections to briefly discuss the two current model and would obtain the thermoelectric coefficients. We shall also admit the assumption about charges made in \cite{seo} and list analytical results for all the thermoelectric coefficients, which may be directly compared with the results of the single current model obtained in section (\href{sec3}{III}). \\  
\noindent We start by considering the following action with two Born-Infeld currents, 
\begin{eqnarray}
\mathcal{S} = \int d^4x \sqrt{-g} \bigg(R+6+\mathcal{L}_{BI_1} + \mathcal{L}_{BI_2}\bigg)
\label{Action2}
\end{eqnarray}
where $$\mathcal{L}_{BI_i} = \dfrac{1}{b_i}\bigg(1-\sqrt{1+\dfrac{b_i}{2}F_i^2}\bigg)~; ~~~~~(i = 1,2)~.$$ 
Here, the Faraday tensors for the currents are given by $F_i = \partial_{[\mu}A_{i\nu]}$, where ($i=1,2$). The equations of motion, in this case, are following,
\begin{eqnarray}
 E_{\mu\nu} \equiv R_{\mu\nu}+3g_{\mu\nu}-\dfrac{1}{2}T_{\mu\nu}^{BI_1} - \dfrac{1}{2}T_{\mu\nu}^{BI_2}=0
\label{EOM_III} \\
\nabla_{\mu}\bigg(\dfrac{F_i^{\mu\nu}}{\sqrt{1+\dfrac{b_i}{2}F_i^2}}\bigg)=0~; ~~(i = 1,2)~.
 \label{EOM_IV}
\end{eqnarray}
In the background geometry given by eq.(\ref{metric}), following a similar analysis as mentioned in section (\href{sec2}{II} and \href{sec3}{III}) with two constant electric field perturbations, namely, $E_1$ and $E_2$ and the thermal perturbation $\zeta$, one may get the following results easily,
\begin{eqnarray}
q_i=\int \Sigma^{(0)} a_{it}^{(0)} \bigg(1+\dfrac{b_i}{2}(a_{it}^{(0)})^2\bigg) +\mathcal{O}(r)~;~~(i = 1,2)~~~
\label{Charge2} \\
\mathcal{Q} = -4\pi T \delta g_{tx}^{(0)}~~~~~~~~~~~~~~~~~~~~~~~~~~~~~~~~~~~~~~~~~~~~~~~~~~
\label{Q_at_leading2} \\
J_i =  \bigg(1+\dfrac{b_i}{2} (a_{it}^{(0)})^2\bigg)(E_i + \partial_x \delta a_{it}^{(0)} - a_{it}^{(0)} \delta g_{tx}^{(0)})~~~~~~~~~~
\label{electric current6} \\
\partial_x(4\pi T\delta g_{tr}^{(0)}) + a_{1t}^{(0)}\bigg(J_1 + a_{1t}^{(0)} \bigg(1+\dfrac{b_1}{2} (a_{1t}^{(0)})^2\bigg)\delta g_{tx}^{(0)}\bigg) ~~\nonumber \\
+ a_{2t}^{(0)}\bigg(J_2 + a_{2t}^{(0)} \bigg(1+\dfrac{b_2}{2} (a_{2t}^{(0)})^2\bigg)\delta g_{tx}^{(0)}\bigg) + 4\pi T \zeta = 0. ~~~
\label{Constraint16} 
\end{eqnarray}
Here $q_i$, $\mathcal{Q}$, and $J_i$ are charges, heat current and electric currents respectively. It should be noted that these expressions are given in terms of the blackhole horizon data only. In the presence of two currents, heat current given by eq.(\ref{Heat Current2}) modifies into the following equation \cite{seo},
\begin{eqnarray}
\mathcal{Q} = U^2\bigg(\partial_r\bigg(\dfrac{\delta g_{tx}}{U}\bigg) - \partial_x\bigg(\dfrac{\delta g_{tr}}{U}\bigg)\bigg) - a_{1t} J_1 - a_{2t} J_2~.~~
\label{Heat Current3}
\end{eqnarray}
Now with some mathematical rearrangements of eqs.(\ref{Q_at_leading2}, \ref{electric current6}, \ref{Constraint16}) and integrations over a period of $x$, one may write the following equations, 
\begin{eqnarray}
J_1 = \dfrac{1}{X_{b_1}}\bigg(1+\dfrac{(\int a_{1t}^{(0)})^2}{YX_{b_1}}\bigg) E_1 + \dfrac{(\int a_{1t}^{(0)})(\int a_{2t}^{(0)})}{YX_{b_1}X_{b_2}} E_2 ~~~~~\nonumber \\
 + \dfrac{(4\pi T\int a_{1t}^{(0)})}{YX_{b_1}}\zeta ~~ \label{two current eq1}\\
 J_2 =  \dfrac{(\int a_{1t}^{(0)})(\int a_{2t}^{(0)})}{YX_{b_1}X_{b_2}} E_1 + \dfrac{1}{X_{b_2}}\bigg(1+\dfrac{(\int a_{2t}^{(0)})^2}{YX_{b_2}}\bigg) E_2~~~~ \nonumber \\
 + \dfrac{(4\pi T\int a_{2t}^{(0)})}{YX_{b_2}} \zeta ~~ \label{two current eq2}\\
 \mathcal{Q}=  \dfrac{(4\pi T) \int a_{1t}^{(0)}}{YX_{b_1}} E_1 +\dfrac{(4\pi T) \int a_{2t}^{(0)}}{YX_{b_2}} E_2 +\dfrac{(4\pi T)^2}{Y} \zeta~~~~
 \label{two current eq3}
\end{eqnarray}
where 
\begin{eqnarray}
X_{b_i} \equiv \int \dfrac{1}{\bigg(1+\dfrac{b_i}{2}(a_{it}^{(0)})^2\bigg)}~;~~~~(i=1,2) ~~~~~~~~~~~~~~~~~~~ \label{X_bi}\\
 Y \equiv \sum_{i=1,2}\bigg\{\int (a_{it}^{(0)})^2 \bigg(1 + \dfrac{b_i}{2}(a_{it}^{(0)})^2\bigg) - \dfrac{(\int a_{it}^{(0)})^2}{X_{b_i}}\bigg\}~.~~
\label{Y}
\end{eqnarray} 
Now we may write the generalized Ohm's law in the presence of two electric perturbations $E_1$, $E_2$, and a thermal perturbation $\zeta$ as $J_i = \Sigma_{ij} E_j$, with the identifications $J_3 \equiv \mathcal{Q}$ and $E_3 \equiv \zeta$, where $\Sigma_{ij}$ is given by the following matrix \cite{seo},
\begin{eqnarray}
\Sigma \coloneqq \begin{pmatrix}
    \sigma_1       & \delta & \alpha_1 T \\
    \bar \delta       & \sigma_2 &\alpha_2 T \\
    \bar \alpha_1 T      & \bar \alpha_2 T & \bar \kappa T 
\end{pmatrix}~.
\label{General_Ohm}
\end{eqnarray}
Using eqs.(\ref{two current eq1}, \ref{two current eq2}, \ref{two current eq3}), we may directly read off elements of the matrix $\Sigma$, which are given below,
\begin{eqnarray}
\sigma_i =  \dfrac{1}{X_{b_i}}\bigg(1+\dfrac{(\int a_{it}^{(0)})^2}{YX_{b_i}}\bigg) \label{response5}\\
\delta = \dfrac{(\int a_{1t}^{(0)})(\int a_{2t}^{(0)})}{YX_{b_1}X_{b_2}} = \bar \delta ~~ \label{response6}\\
\alpha_i = \dfrac{4\pi \int a_{it}^{(0)}}{YX_{b_i}} = \bar \alpha_i ~~~~~~~~ \label{response7} \\
\bar \kappa = \dfrac{(4\pi )^2 T}{Y}~.~~~~~~~~~~~~~~~ \label{response8}
\end{eqnarray}
Here, $\sigma_1$ and $\sigma_2$ are the electric conductivities and $\bar \kappa$ is the heat conductivity, while $\delta$, $\bar \delta$, $\alpha_i$, and $\bar \alpha_i$ are thermoelectric conductivities. It should be noted that $\delta = \bar \delta$ and $\alpha_i = \bar \alpha_i$, which is the Onsager relation in the presence of two currents. The thermal heat conductivity, defined when all other currents vanish, could be obtained by setting $J_1 = 0$ and $J_2 = 0$ in eqs.(\ref{two current eq1}, \ref{two current eq2}),  
\begin{eqnarray}
\kappa \equiv \dfrac{\mathcal{Q}}{T\zeta}\bigg|_{J_1 = 0 = J_2} = \dfrac{(4\pi)^2 T}{\sum_{i=1,2}\bigg\{\int (a_{it}^{(0)})^2 \bigg(1+ \dfrac{b_i}{2}(a_{it}^{(0)})^2\bigg)\bigg\}} ~.~~
\label{response9}
\end{eqnarray}
It has been mentioned earlier that the two current holographic model proposed in \cite{seo} explained the experimental results under the assumption that the two charges $q_1$ and $q_2$ are proportional to each other. Hence, in this case, we now make the following ansatz,
\begin{eqnarray}
a_{2t}^{(0)} =  g a_{1t}^{(0)} 
\label{ansatz1} \\
b_2 = \dfrac{b_1}{g^2}~~~~
\label{ansatz2}
\end{eqnarray}
which would imply that $q_2 = g q_1$. With these ansatzs, the thermoelectric DC response obtained for the two current model simplifies to the following results,
\begin{eqnarray}
  \sigma_1 =  \dfrac{1}{X_{b_1}}\bigg(1+\dfrac{(\int a_{1t}^{(0)})^2}{(1+g^2)X}\bigg) ~~~~~~~~~~~~~~~~~~~\label{response9} \nonumber \\
  \sigma_2 =  \dfrac{1}{X_{b_1}}\bigg(1+\dfrac{g^2(\int a_{1t}^{(0)})^2}{(1+g^2)X}\bigg) ~~~~~~~~~~~~~~~~~~\label{response10}\nonumber \\
  \delta = \dfrac{g(\int a_{1t}^{(0)})^2}{(1+g^2)X X_{b_1}} = \bar \delta ~~~~~~~~~~~~~~~~~~~~~~~~~ \label{response11}\nonumber \\
  \alpha_1 = \dfrac{4\pi \int a_{1t}^{(0)}}{(1+g^2)X} = \bar \alpha_1 ~~~~~~~~~~~~~~~~~~~~~~~~~~~ \label{response12} \nonumber   \\
  \alpha_2 = g \alpha_1 = \bar \alpha_2 ~~~~~~~~~~~~~~~~~~~~~~~~~~~~~~~~~~~~ \label{response13} \nonumber
  \end{eqnarray}
\begin{eqnarray}
\bar \kappa = \dfrac{(4\pi )^2 T X_{b_1}}{(1+g^2)X}~~~~~~~~~~~~~~~~~~~~~~~~~~~~~~~~~~~~ \label{response14}  \nonumber\\
\kappa = \dfrac{(4\pi)^2 T}{(1+g^2)\bigg\{\int (a_{1t}^{(0)})^2 \bigg(1+ \dfrac{b_1}{2}(a_{1t}^{(0)})^2\bigg)\bigg\}} ~.~
\label{response15}
\end{eqnarray}
Also it should be noted that under the considered ansatz given by eqs.(\ref{ansatz1}, \ref{ansatz2}), $X_{b_1} = X_{b_2} $. Here, $X$ is given by, 
$$X = \bigg[\int (a_{1t}^{(0)})^2 \bigg(1+\dfrac{b_1}{2} (a_{1t}^{(0)})^2\bigg)\bigg] X_{b_1} - \bigg(\int a_{1t}^{(0)}\bigg)^2~.$$
The choice of notation $X$, in eq.(\ref{response15}), is being made to emphasize that these results could directly be compared with the results of the single current model given in sections (\href{sec2}{II} and \href{sec3}{III}). Although, there are no results available in the literature for the two current model for inhomogeneous holographic lattice with Maxwell electrodynamics, we may obtain these by simply taking the limit $b_1 \rightarrow 0$ in the above results.

\section{Conclusion and Remarks}
\noindent In this work, we have obtained analytical expressions for the thermoelectric DC response for an inhomogeneous holographic lattice model with Born-Infeld corrections. To have inhomogeneous holographic lattice, we have used spatially modulated chemical potential in the $x$-direction as proposed in some of the earlier works \cite{CLS, GTHJES, Donos6}. We have observed that the presence of the Born-Infeld currents changes the thermoelectric response non-trivially, which can be expressed as a correction to the various conductivity expressions with Born-Infeld parameter, $b$. With the limit $b \rightarrow 0$, one may recover conductivity expressions for the Maxwell case. We have then used these results to built a two current holographic model for Graphene, near the Dirac point, in the spirit of \cite{seo, Rogatko1,Rogatko2}. However, we have considered two independent Born-Infeld currents in this model along with an inhomogeneous holographic lattice, which has not been studied before. We have obtained analytic expressions for various conductivities with the assumption that the charges associated with both currents are proportional to each other. Again, results for the two Maxwell current case may be readily obtained by taking the limits $b_1 \rightarrow 0$ and $b_2 \rightarrow 0$. It should also be noted that low and high temperature limits for this model is easy to obtain from the general results mentioned in the previous sections.  

\noindent {\bf{Acknowledgements}}:  Ankur Srivastav would like to acknowledge Aristomenis Donos for a communication regarding this work. Authors would also like to thank anonymous referee for fruitful comments.

\end{document}